\documentclass[10pt,twocolumn,letterpaper]{article}

\usepackage{cvpr}              

\usepackage[table]{xcolor}
\usepackage{graphicx}
\usepackage{amsmath}
\usepackage{amssymb}
\usepackage{booktabs}
\usepackage{times}
\usepackage{epsfig}
\usepackage{tabularx}
\usepackage{arydshln}
\usepackage{multirow}
\usepackage{array}
\usepackage{latexsym}
\usepackage{CJK}
\usepackage{hhline}
\usepackage{makecell}
\usepackage{algorithm} 
\usepackage{algpseudocode}
\usepackage{bbding}

\usepackage[pagebackref,breaklinks,colorlinks]{hyperref}

\usepackage[capitalize]{cleveref}
\crefname{section}{Sec.}{Secs.}
\Crefname{section}{Section}{Sections}
\Crefname{table}{Table}{Tables}
\crefname{table}{Table}{Tables}

\def\cvprPaperID{6948} 
\def\confName{CVPR}
\def\confYear{2022}

\begin{document}


\title{Learning Pixel-Adaptive Weights for Portrait Photo Retouching}

\author{Binglu\ Wang \ \ \ \ Chengzhe\ Lu*\ \ \ \ Dawei\ Yan*\ \ \ \ Yongqiang\ Zhao\\
*\ Equal Contribution\\
}
\maketitle

\begin{abstract}
Portrait photo retouching is a photo retouching task that emphasizes human-region priority and group-level consistency. The lookup table-based method achieves promising retouching performance by learning image-adaptive weights to combine 3-dimensional lookup tables (3D LUTs) and conducting pixel-to-pixel color transformation. However, this paradigm ignores the local context cues and applies the same transformation to portrait pixels and background pixels when they exhibit the same raw RGB values. In contrast, an expert usually conducts different operations to adjust the color temperatures and tones of portrait regions and background regions. This inspires us to model local context cues to improve the retouching quality explicitly. Firstly, we consider an image patch and predict pixel-adaptive lookup table weights to precisely retouch the center pixel. Secondly, as neighboring pixels exhibit different affinities to the center pixel, we estimate a local attention mask to modulate the influence of neighboring pixels. Thirdly, the quality of the local attention mask can be further improved by applying supervision, which is based on the affinity map calculated by the groundtruth portrait mask. As for group-level consistency, we propose to directly constrain the variance of mean color components in the Lab space. Extensive experiments on PPR10K dataset verify the effectiveness of our method, \eg on high-resolution photos, the PSNR metric receives over 0.5 gains while the group-level consistency metric obtains at least 2.1 decreases.
\end{abstract}
\vspace{-0.2cm}
\vspace{-0.2cm}
\section{Introduction}
\label{sec:intro}

\begin{figure}[htbp]
\centering
\includegraphics[width=01\linewidth]{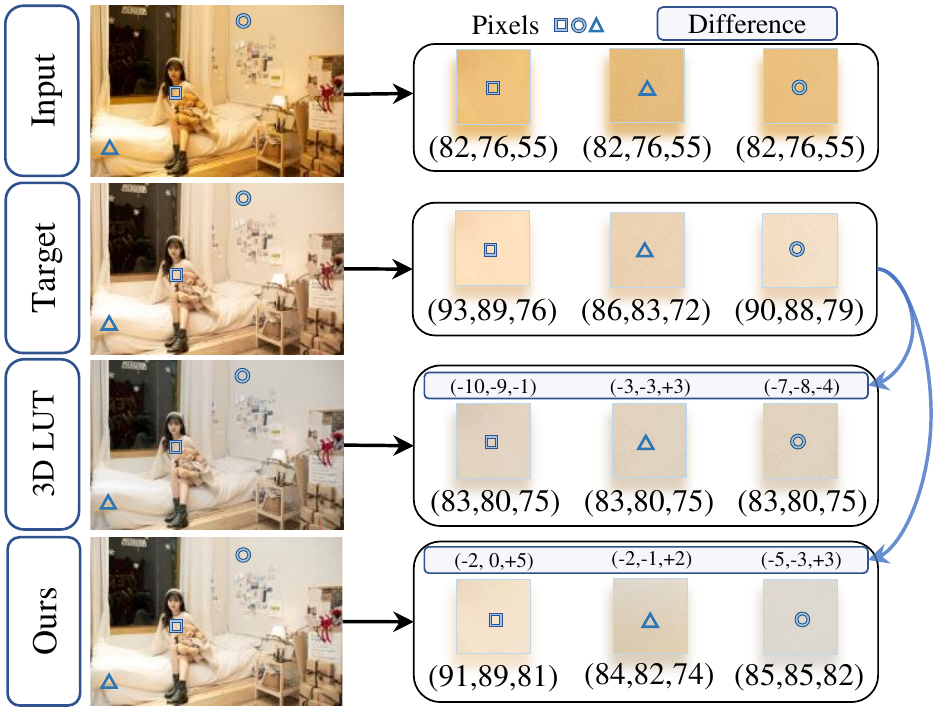}
\caption{Comparison of retouching quality. The outputs of 3D LUT are identical for the same RGB inputs, but the outputs of our method vary according to context pixels. The RGB values of each pixel are shown in parentheses.}
\label{FigMotivation}
\vspace{-0.3cm}
\end{figure}

With promising development, the image enhancement community has observed significant advances and received numerous well-performed methods \cite{huang2018range, moran2021curl, zeng2020learning, vu2018fast, zhang2017beyond, tai2017memnet, zhang2020residual}. Recently, Liang \etal focused on the visual quality of portrait photos and proposed the portrait photo retouching task. When recording meaningful moments by a photograph, the human often acts as an essential character in the photo, and people usually take multiple photos in similar scenes. However, the quality of raw photos taken directly by a cell phone or camera is easily affected by weather, lighting and shooting equipment, \etc. Meanwhile, the tone of multiple photos taken on the same subject at the same scene may also become different. Liang \etal \cite{liang2021ppr10k} indicated two critical factors for portrait photo retouching: one is the human-region priority, \ie the human region in the photo should receive more refined retouching compared to the background region. The other is group-level consistency,~\ie a group of portrait photos should be retouched to a consistent tone, even though the shooting conditions are different.

The lookup table-based method \cite{zeng2020learning} exhibits significant performance for portrait photo retouching. In particular, Zeng \etal \cite{zeng2020learning} proposed an image-adaptive 3D LUT method that combines multiple lookup tables with a convolutional neural network. Recently, based on 3D LUT, Liang \etal \cite{liang2021ppr10k} weighted the MSE loss to emphasize the human region and build a baseline method for portrait photo retouching. Actually, in the photo retouching process, experts usually perform different operations on portrait and background regions. As shown in \cref{FigMotivation}, even raw pixels exhibit the same RGB values, and their retouching targets could be different because each pixel has its specific contexts. However, the current method \cite{liang2021ppr10k} tackles the whole image and only generates the image-adaptive weights to combine the looking-up results. In contrast, when contextual information is considered, our proposed method can adaptively adjust the retouching results and achieve high-quality enhancement.

An intuitive way to consider the contextual information of a pixel is to select an image patch containing $k \times k$ pixels and predict pixel-adaptive weights to combine the looking-up results. 
Besides, in an image patch, the contextual pixels may exhibit different affinities with respect to the center pixel. Thus, we propose to enlarge the influence of homogeneous pixels (\eg two portrait pixels) and suppress the influence of heterogeneous pixels (\eg a portrait pixel and a background pixel) via learning a pixel-adaptive local attention mask. Moreover, considering each photo is accompanied by a portrait mask, we can construct the groundtruth affinity maps and apply pixel-wise supervision, which can further improve the quality of local attention masks. The above processes form the \textit{Local-context Aware Module} (LAM). It explores contextual cues, predicts pixel-adaptive weights, and brings complementary information to image-adaptive LUT weights. Consequently, our proposed method can effectively integrate lookup table results and achieve high-quality retouching results.

In addition to the retouching of individual photos, ensuring the consistent color temperatures and tones of a group of photos is also important for the portrait photo retouching task. Liang \etal \cite{liang2021ppr10k} simulate the group-level variations by cropping two patches from the same image and applying different transformations. We propose a group-style aware module, which is simple to implement but can effectively guide a group of photos to exhibit consistent color temperatures and tones. In particular, given a group of photos, we first convert the enhanced photo and all other target photos to the Lab space and then constrain the variance of mean color components to achieve the group-style aware retouching.

The contributions of this paper are summarized as follows:
\begin{itemize}
\vspace{-0.2cm}
\item We reveal the influence of contextual cues to the portrait photo retouching task and design the local-context aware module to explicitly consider context pixels and predict pixel-adaptive weight.
\vspace{-0.2cm}
\item We propose the group-style aware module, which performs in a simple manner but can effectively ensure consistent color temperatures and tones among a group of enhanced photos. 
\vspace{-0.2cm}
\item Extensive experiments on the large-scale PPR dataset demonstrate the efficacy of our proposed method, \eg given high-resolution portrait photos, the PSNR metric receives over 0.5 gains, and the group-level consistency metric obtains at least 2.1 decreases.
\end{itemize}


\begin{figure*}[thbp]
\centering
\includegraphics[width=1\linewidth]{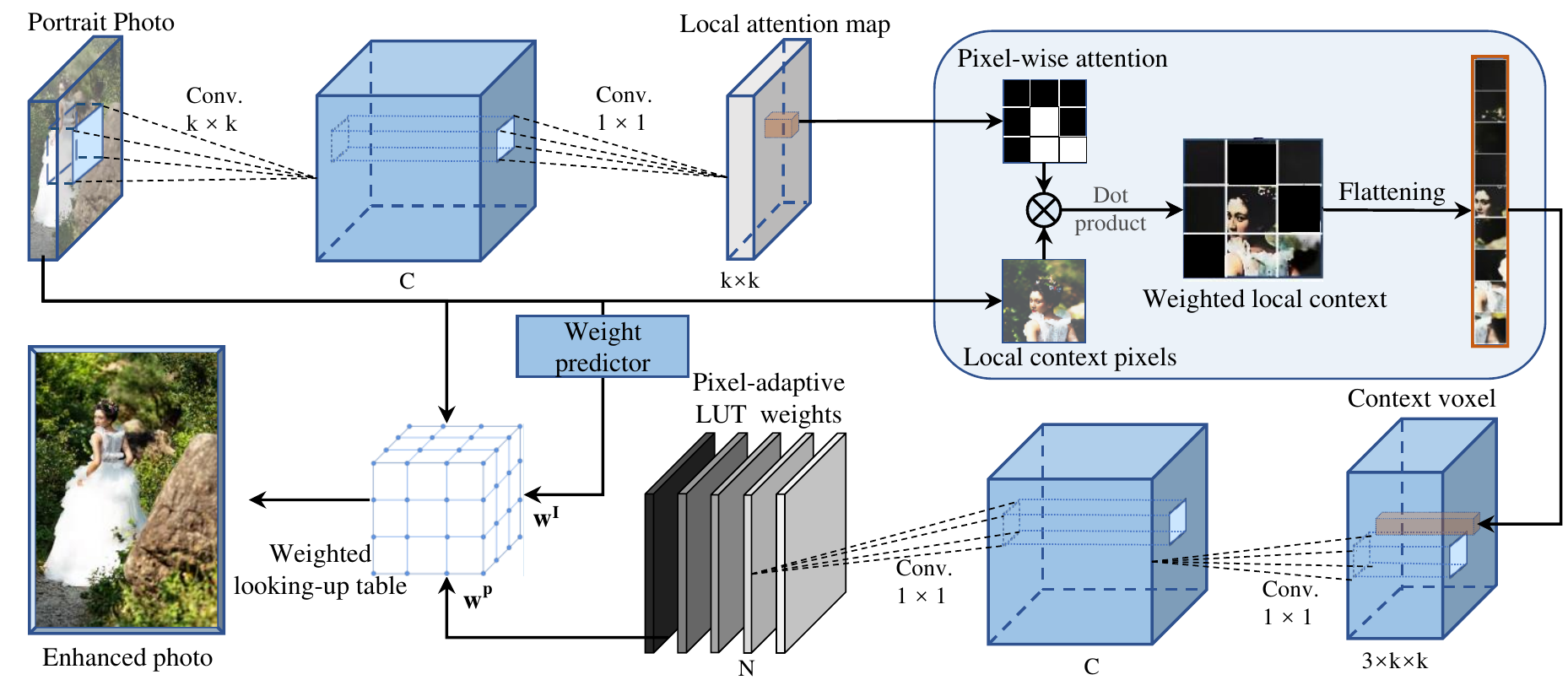}
\caption{Framework of the proposed local-context aware module. Given a portrait photo, we first elaborately estimate pixel-wise attention and then adaptively modulate local context pixels. Afterward, neighboring pixels are flattened to form the context voxel, which are used to predict pixel-adaptive LUT weights $\mathbf{w}^{p}$. Meanwhile, a weight predictor estimates the image-adaptive weights $\mathbf{w}^{I}$. Finally, the weighted looking-up table operation generates the enhanced photo.}
\label{bigfigureb}
\end{figure*}

\section{Related work}
\textbf{Learning-based image restoration. }
Image restoration \cite{cai2018learning, hasinoff2016burst, wang2019underexposed, liu2021pd, zamir2021multi, fu2021auto, zheng2021ultra} aims to remove image distortions caused by various situations. Since manual retouching requires a mass of expert knowledge and subjective judgment, many learning-based image restoration methods come out recently. For image exposure and low-light issues, Wang \etal \cite{wang2019underexposed} propose a neural network for photo enhancement using deep lighting estimation to fix underexposed photos. Recently Mahmoud \etal \cite{afifi2021learning} propose a model that can perform multi-scale exposure correction for images. Guo \etal \cite{Zero-DCE} propose a model based on a luminance enhancement curve, which is iterated continuously to achieve gradual contrast and luminance enhancement of low-light images. As for the noise reduction task, Huang \etal \cite{Huang_2021_CVPR} present a model which generates two independent noise-bearing images of similar scenes to train the noise reduction network.

\textbf{Learning-based image retouching. }
Image retouching aims to improve the quality of raw images and has received a large number of significant advances \cite{chen2018deep, gharbi2017deep, he2020conditional, ignatov2017dslr, kim2020global, kim2020pienet, moran2020deeplpf, park2018distort, song2021starenhancer, zhang2021star, kim2021representative}. Focusing on real-time processing of high-resolution images (\eg over 24 megapixels), Gharbi \etal proposed the HDRNet\cite{gharbi2017deep} that put most of the computation on downsampled images. He \etal \cite{he2020conditional} proposed the CSRNet that extracted global features with normal CNN and modulated the features after 1$\times$1 convolution. Some researchers, such as Sean \etal \cite{moran2020deeplpf} and Han-Ul \etal \cite{kim2020global}, adjusted the research paradigm and introduced residuals into the model in order to pursue an enhancement over the original image instead of training the image directly. In addition to expert retouching, Han-Ul \etal \cite{kim2020pienet} developed PieNet to solve the problem of personalization in image enhancement. These excellent works demonstrate that image enhancement has splendid results based on paired datasets. However, the collection of paired datasets is expensive. Therefore, image enhancement methods based on unpaired datasets started to receive attention. There are some GAN-based methods \cite{chen2018deep, ignatov2017dslr}, while do not require pairs of input images and target images. Meanwhile, Jongchan \etal \cite{park2018distort} proposed the Distort-and-Recover method focusing on image color enhancement, which also only required high-quality reference images for training. Although existing image enhancement models have good generalizability, our work focuses on portrait photo retouching, which emphasizes both the human-region priority and group-level consistency, leaving it an under-explored task.

In addition, HDRNet \cite{gharbi2017deep} can be regarded as a masterpiece along the lines of bilateral filter. For image processing, the features extracted by the network are complete. The applicability of this method is very broad, considering that the human photographer's retouching process also considers only these features. CSRNet \cite{he2020conditional} used an end-to-end approach that is easy to train. By utilizing an expanded convolutional layer, CSRNet can extend the field of perception without the loss of the receptive field.

\textbf{Deep learning methods with 3D LUT. }
In previous studies, Zeng \etal \cite{zeng2020learning} combined the deep learning-based weight predictor and 3D LUT for the first time, which achieved outstanding results. In particular, 3D LUT learned a weight predictor based on the convolutional network, which tackled low-resolution images and estimated image-adaptive weights. Then, multiple lookup tables were dynamically combined to capture image characteristics and perform the enhancement. The 3D LUT was a representative work that combined the deep learning paradigm with the traditional image enhancement paradigm, which presented the community with a new perspective on how to implement image enhancement. Subsequently, Jo \etal \cite{jo2021practical} implemented a similar lookup process from input to output values by training a deep super-resolution network and transferring the output values of the deep learned model into the LUT. 
A lot of works show that the traditional LUTs can be well integrated with image enhancement tasks. In addition, due to the hardware-friendly nature, LUTs have been used in camera imaging pipelines to retrieve precomputed output from memory.

\textbf{Contextual information in image enhancement. }
Contextual information can be taken into account in image enhancement to get better results. Its importance has been demonstrated by a number of distinguished researchers in recent work. By effectively modeling contextual information, many deep learning-based methods \cite{li2018recurrent, yin2018fisheyerecnet, chen2021canet} were able to remove rain marks, distortions, and shadows from images. Zhang \etal \cite{zhang2021context} proposed a context reasoning attention network to adaptively modulate the convolution kernel according to the global context and apply it to super-resolution tasks. Besides, there was also some work that applied contextual information to complete a missing region \cite{zeng2019learning, zeng2021cr}. The context information also plays an important role on other computer vision domains, \eg image classification \cite{tang2015improving}, saliency detection \cite{liu2018picanet}. In particular, Liu \etal \cite{liu2018picanet} proposed PiCANet to learn informative contextual features of each pixel and then embed them into a saliency detection network. In the photo retouching domain, the contextual information is still under-explored. In this work, we make an early exploration to study the influence of contextual information on the portrait photo retouching task and observe sufficient improvements.

\textbf{Spatial-aware 3D LUT. }
Spatial-aware 3D LUT \cite{wang2021real} is a most relevant work to ours, though our work is independently developed with \cite{wang2021real}. The insight of learning pixel-adaptive weights is similar. However, we precisely model local context cues by a convolutional layer with kernel size $k \times k$, while the UNet architecture in \cite{wang2021real} considers the entire image and is affected by both local contexts and global information. As a result, in \cite{wang2021real}, the global information in the UNet branch is somewhat redundant with the image-adaptive weights, also has the risk of bringing noises to the retouching process. Besides, \cite{wang2021real} employed the UNet architecture and maintained multiple lookup tables. It is more complex than our local-context aware module.

\section{Methodology}



\subsection{Image enhancement based on 3D LUT}
\label{sec:3D LUT}

3D LUT defines a 3D grid consisting of $M^{3}$ elements, where $M$ indicates the number of bins in each color channel. Given an input image $I$, the looking-up table method makes transformations and generates the enhanced output $O$. The transformation operation can be formulated as:
\begin{equation}
\label{eq:3D LUTlookup}
    O_{(i,j,k)}=\varphi(I_{(i,j,k)}^{R},I_{(i,j,k)}^{G},I_{(i,j,k)}^{B}),
\end{equation}
where $\varphi(\cdot)$ defines a pixel-to-pixel mapping via a looking-up table, in practice, $M$=33 is a common setting, which contains 108K parameters and achieves a good balance between inference speed and enhancement quality.

To tackle dramatic variations among multiple photos, Zeng \etal \cite{zeng2020learning} capture the holistic cues within an image and predict image-adaptive weight $\textbf{w}^{I} = [\textbf{w}^{I}_{1}, \textbf{w}^{I}_{2}\dots \textbf{w}^{I}_{n}]$ to adjust the looking-up table results. Thus, the transformation operation with LUT weight can be formulated as follows:
\begin{equation}
\label{eq:3D LUTlookupaddweight}
\begin{aligned}
    O_{(i,j,k)}&=\Phi(I_{(i,j,k)}^{R},I_{(i,j,k)}^{G},I_{(i,j,k)}^{B},\textbf{w}^{I}_{1}\dots \textbf{w}^{I}_{n})\\
    &=(\sum_{n=1}^{N} \textbf{w}^{I}_{n}\times\varphi_{n})(I_{(i,j,k)}^{R},I_{(i,j,k)}^{G},I_{(i,j,k)}^{B}),
\end{aligned}
\end{equation}
where $\Phi(\cdot)$ indicates the image enhancement process and $\varphi_{n}$ is the mapping operation of the $n^{th}$ looking-up table.


\subsection{Local-context aware module}
\label{sec:LC}

In the portrait photo retouching task, experts usually apply different adjustments about color temperatures and tones for human regions and backgrounds. Thus, the context cues should be carefully considered in the retouching process.


As shown in \cref{bigfigureb}, we propose the local-context aware module to predict pixel-adaptive weights and precisely modulate the looking-up table results. Given a portrait photo $\mathbf{x}$, each pixel considers its $k \times k$ neighboring pixels to perceive local context cues. To this end, we first employ a convolutional layer with kernel size $k \times k$ to extract context features $\mathbf{f}_{\rm ctx} = {\rm ReLU}(\mathbb{C}^{k \times k} (\mathbf{x}))$. Then, we predict local attention map $\mathbf{a} \in \mathbb{R}^{(k \times k) \times H \times W}$:
\begin{equation}
    \mathbf{a} = {\rm Softmax}(\mathbb{C}^{1 \times 1} (\mathbf{\mathbf{f}_{\rm ctx}})),
\end{equation}
where ``${\rm Softmax}$" indicates softmax normalization among $k^{2}$ attention elements. $\mathbb{C}^{k \times k}(\cdot)$ means a ${k \times k}$ convolutional layer.

\begin{figure}[thbp]
\centering
\includegraphics[width=0.45\textwidth]{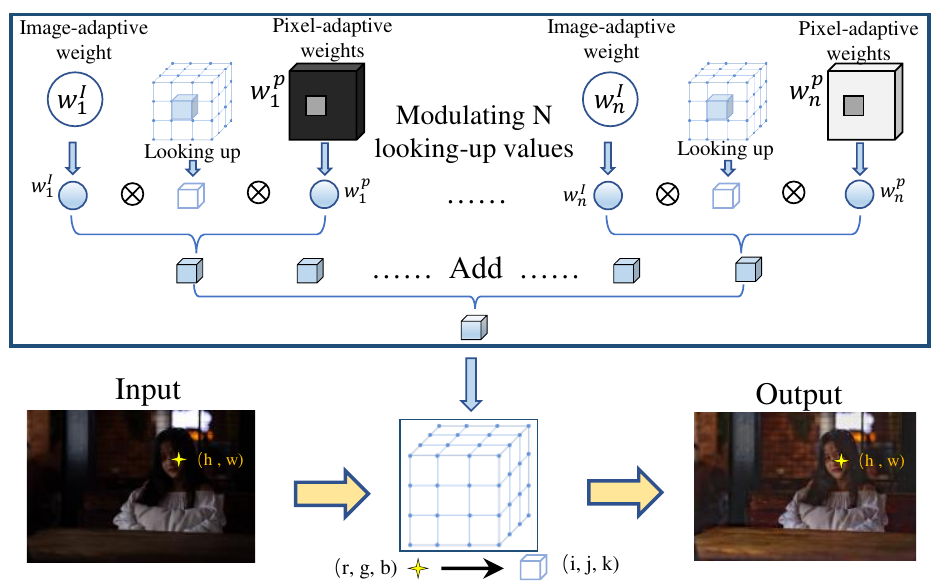}
\caption{Portrait photo retouching process with pixel-adaptive weights and image-adaptive weights. The lookup table generates a retouching result for a pixel, which is then jointly modulated by pixel-adaptive weight ${\textbf{w}}^{p}$ and image-adaptive weight $\textbf{w}^{I}$. Finally, multiple retouching results are added to enhance the input photo.}
\label{LUTintegration}
\end{figure}

Considering a pixel $x_{i,j}$, the pixel-wise attention weight $\mathbf{a}_{i,j}$ can be used to modulate its neighboring context pixels via dot product. Then, similar context pixels would be strengthened while the influence of intrusive pixels would be weakened. Next, all neighboring pixels are flattened to a vector, and context vectors from all spatial locations form the neighboring context voxel $\mathbf{v} \in \mathbb{R}^{(3\times k \times k) \times H \times W}$. Afterwards, we employ a convolutional layer to tackle the context voxel and obtain feature $\mathbf{f}_{\rm vox} = {\rm ReLU}(\mathbb{C}^{1 \times 1} (\mathbf{v}))$. In the end, a convolutional layer with kernel size 1 can adaptively predict LUT weights for each pixel:
\begin{equation}
    \mathbf{w}^{p} = \mathbb{C}^{1 \times 1} (\mathbf{\mathbf{f}_{\rm vox}}).
\end{equation}

At each spatial location, the pixel-adaptive weights ${\textbf{w}}^{p}$ contains $N$ elements, being responsible for zooming or shrinking the looking-up table results.


\subsection{Fuse the multiple 3D LUTs.}
\label{sec:LUT}

\textbf{Image-adaptive Weight. }Given $N$ 3D LUTs, the fusion is performed according to the weight $\textbf{w}^{I}$ output by the image classifier. In our work, the number of 3D LUTs is set to 5 based on ablation experiments.

\textbf{Pixel-adaptive Weight. }The fuse effect based on classifier and 3D LUTs is useful, but it can be further improved by considering the contextual information of each pixel. In this paper, we consider the contextual information for each pixel and estimate pixel-adaptive attention weights $\mathbf{w}^{p}$. By jointly considering image-adaptive attention weights $\mathbf{w}^{I}$ and pixel-adaptive attention weights  $\mathbf{w}^{p}$, our photo retouching method can perceive both the image-level holistic information and pixel-level contextual cues and achieve high-quality image enhancement.

\cref{LUTintegration} illustrates the fuse process by tackling a certain pixel located at (h,w) within the photo. The program sends the photo to a group of 3D LUTs and obtains the converted result. The result is then integrated with the $\mathbf{w}^{I}$ and $\mathbf{w}^{p}$ to obtain the RGB value at the corresponding pixel point (h, w) of the output photo. This procedure is performed for each pixel, which generates the complete enhanced photo. 

\begin{figure}[thbp]
\centering
\includegraphics[width=1\linewidth]{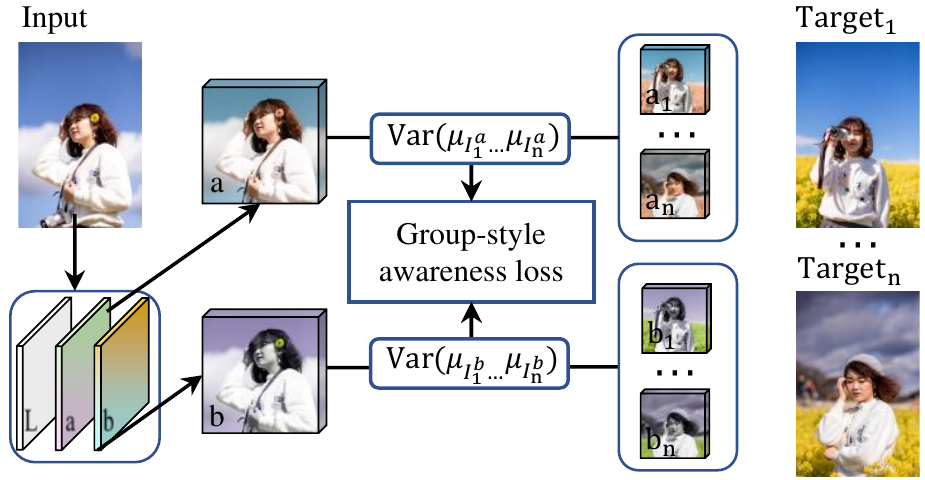}
\caption{Calculation process of the group-style aware loss. The retouched photo and all other target photos from the same group are first converted to Lab space. Then, we calculate the variance of mean color components in a-channel and b-channel to obtain the group-style aware loss.}
\vspace{-0.4cm}
\label{GLCloss}
\end{figure}

\subsection{Group-style aware module}
\label{sec:GAM}
The group-level consistency characteristic emphasizes more the uniformity of style among a group of photos rather than the retouching effect of a single photo. In \cite{liang2021ppr10k}, given a group of enhanced image, Liang \etal \cite{liang2021ppr10k} first convert each image from RGB to Lab color space, and then select the $a$ and $b$ channels to measure the variance of mean color components. Based on the combination of $a$ and $b$ channels, Liang \etal \cite{liang2021ppr10k} defines the GLC metric.

The group-style aware module strategy is to enumerate the color of all pixels of the photo and calculate the mean value $\mu$, calculate the variance between $\mu$ and the value of the same group of target photos. The smaller the variance, the closer the values in the two channels, and the more consistent the color temperatures and tones of the image.

As shown in \cref{GLCloss}, given a group of retouched photos $[\hat{\mathbf{p}}_{1}, \hat{\mathbf{p}}_{2}, ..., \hat{\mathbf{p}}_{n}]$, we first calculate their mean values from the a-channel and b-channel, and then do the same for images generated by the network model retouching. Calculating the variance of these values can further obtain group-style aware loss.

Given a group of photos, we find it is inconvenient and inefficient to simultaneously forward all photos through the network and calculate the group-style aware loss $\mathcal{L}_{GAM}$, because each group contains a different number of photos. Instead, we propose to compare a retouched photo with all other target photos from the same group, where the mean value of a-channel and b-channel can be calculated in advance.


\subsection{Loss function}
\label{sec:Loss}

\begin{figure}[thbp]
\centering
\includegraphics[width=1\linewidth]{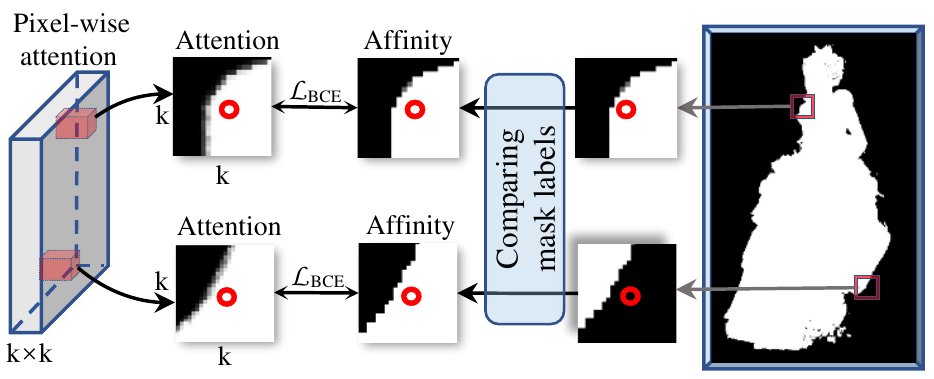}
\caption{Calculation process of the edge supervision loss. Considering a pixel seated on the portrait border area, we compare the mask labels between the center pixel and the neighboring pixels, obtain the affinity map. Afterward, the affinity map can guide the pixel-wise attention map via calculating the BCELoss $\mathcal{L}_{BCE}$.}
\vspace{-0.4cm}
\label{bigfigurea}
\end{figure}

\begin{table*}[thbp]
  \centering
    \caption{Comparison of portrait photo retouching results on PPR10K dataset. a/b/c indicate the dataset retouched by three experts. The ``LR" column and the ``HR" column report the test results of each item on 360P and 4K $\sim$ 8K photos, respectively.}
    \begin{tabular}{rccccccccccc}
    \toprule
    \multirow{2}[2]{*}{Method} & \multirow{2}[2]{*}{Dataset} & \multicolumn{2}{c}{$PSNR \uparrow$} & \multicolumn{2}{c}{$\Delta E_{ab} \downarrow$} & \multicolumn{2}{c}{$PSNR^{HC}\uparrow$} & \multicolumn{2}{c}{$\Delta E_{ab}^{HC}\downarrow$} & \multicolumn{2}{c}{$M_{GLC}\downarrow$} \\
             &          & LR       & HR       & LR       & HR       & LR       & HR       & LR       & HR       & LR       & HR \\
    \hline
    \multicolumn{1}{r}{HDRNet\cite{gharbi2017deep}} & PPR10K-a & 23.93    & 23.06    & 8.70     & 9.13     & 27.21    & 26.58    & 5.65     & 5.84     & 14.83    & 14.37  \\
    \multicolumn{1}{r}{CSRNet\cite{he2020conditional}} & PPR10K-a & 22.72    & 22.01    & 9.75     & 10.20    & 25.90    & 25.19    & 6.33     & 6.73     & 12.73    & 12.66  \\
    \multicolumn{1}{r}{3D LUT\cite{zeng2020learning}} & PPR10K-a & 25.64    & 25.15    & 6.97     & 7.25     & \textbf{28.89} & 28.39    & 4.53     & 4.71     & 11.47    & 11.05  \\
    \multicolumn{1}{r}{Liang \etal \cite{liang2021ppr10k}} & PPR10K-a & \textbf{25.99} & \textbf{25.55} & \textbf{6.76} & \textbf{7.02} & 28.29    & \textbf{28.83} & \textbf{4.38} & \textbf{4.55} & 10.81    & 10.32  \\
    \multicolumn{1}{r}{Liang \etal \cite{liang2021ppr10k}+GLC} & PPR10K-a & 25.31    & 24.60    & 7.30     & 7.75     & 28.56    & 27.86    & 4.75     & 5.03     & \textbf{9.95} & \textbf{9.68} \\
    \multicolumn{1}{r}{Ours (LAM)}     & PPR10K-a & \textbf{26.23} & \textbf{26.14}          & \textbf{6.62} &\textbf{6.66}          & \textbf{29.53} &\textbf{29.41}          & \textbf{4.29} &\textbf{4.32}          & 10.77  &10.71  \\
    \multicolumn{1}{r}{Ours (LAM+GAM)}     & PPR10K-a & 25.66 & 25.57 & 7.42  & 7.47 & 28.96 & 28.85 & 4.82 & 4.85  & \textbf{6.67} &\textbf{6.66} \\
    \hline
    \multicolumn{1}{r}{HDRNet\cite{gharbi2017deep}} & PPR10K-b & 23.96    & 23.51    & 8.84     & 9.13     & 27.21    & 26.55    & 5.74     & 5.92     & 13.21    & 13.04  \\
    \multicolumn{1}{r}{CSRNet\cite{he2020conditional}} & PPR10K-b & 23.76    & 23.29    & 8.77     & 9.28     & 27.01    & 26.62    & 5.68     & 5.90     & 11.82    & 11.73  \\
    \multicolumn{1}{r}{3D LUT\cite{zeng2020learning}} & PPR10K-b & 24.70    & 24.30    & 7.71     & 7.97     & 27.99    & 27.59    & 4.99     & 5.16     & 9.90     & 9.52  \\
    \multicolumn{1}{r}{Liang \etal \cite{liang2021ppr10k}} & PPR10K-b & \textbf{25.06} & \textbf{24.66} & \textbf{7.51} & \textbf{7.73} & \textbf{28.36} & \textbf{27.93} & \textbf{4.85} & \textbf{5.00} & 9.87     & 9.60  \\
    \multicolumn{1}{r}{Liang \etal \cite{liang2021ppr10k}+GLC} & PPR10K-b & 24.52    & 23.81    & 7.93     & 8.42     & 27.82    & 27.12    & 5.12     & 5.44     & \textbf{9.01} & \textbf{8.73} \\
    \multicolumn{1}{r}{Ours (LAM)} & PPR10K-b & \textbf{25.35} & \textbf{25.28} & \textbf{7.31} & \textbf{7.34} & \textbf{28.63} & \textbf{28.55} & \textbf{4.73} & \textbf{4.75} & 9.29 & 9.25 \\
    \multicolumn{1}{r}{Ours (LAM+GAM)} & PPR10K-b & 24.97 & 24.90 & 8.10 & 8.13 & 28.27 & 28.19 & 5.24 & 5.26 & \textbf{6.56} & \textbf{6.57} \\
    \hline

    \multicolumn{1}{r}{HDRNet\cite{gharbi2017deep}} & PPR10K-c & 24.08    & 23.66    & 8.87     & 9.05     & 27.32    & 26.93    & 5.76     & 5.99     & 14.76    & 14.28  \\
    \multicolumn{1}{r}{CSRNet\cite{he2020conditional}} & PPR10K-c & 23.17    & 22.85    & 9.45     & 9.87     & 26.47    & 26.09    & 6.12     & 6.54     & 14.64    & 14.22  \\
    \multicolumn{1}{r}{3D LUT\cite{zeng2020learning}} & PPR10K-c & 25.18    & 24.78    & 7.58     & 7.85     & 28.49    & 28.09    & 4.92     & 5.09     & 13.51    & 13.16  \\
    \multicolumn{1}{r}{Liang \etal \cite{liang2021ppr10k}} & PPR10K-c & \textbf{25.46} & \textbf{25.05} & \textbf{7.43} & \textbf{7.69} & \textbf{28.80} & \textbf{28.38} & \textbf{4.82} & \textbf{4.98} & 13.49    & 13.06  \\
    \multicolumn{1}{r}{Liang \etal \cite{liang2021ppr10k}+GLC} & PPR10K-c & 24.59    & 24.01    & 8.02     & 8.39     & 27.92    & 27.33    & 5.20     & 5.43     & \textbf{12.76} & \textbf{12.79}  \\
    \multicolumn{1}{r}{Ours (LAM)}     & PPR10K-c & \textbf{25.65} & \textbf{25.56}  & \textbf{7.39} & \textbf{7.43} & \textbf{28.95} & \textbf{28.84} & \textbf{4.80} & \textbf{4.83} & 14.62 & 14.56 \\
    \multicolumn{1}{r}{Ours (LAM+GAM)}     & PPR10K-c & 25.31 & 25.22 & 8.00 & 8.03 & 28.61 & 28.51 & 5.19 & 5.21 & \textbf{8.87} &\textbf{8.84} \\
    \bottomrule
    \end{tabular}%
    \vspace{-0.2cm}
  \label{tab:bigtable}%
\end{table*}%

As we adopt the lookup table paradigm to retouch photos, we follow \cite{zeng2020learning}, adopt the MSE loss $\mathcal{L}_{mse}$, the smooth regularization loss $\mathcal{R}_{s}$ and monotonicity regularization loss $\mathcal{R}_{m}$ as the basic loss terms, which can be calculated as:
\begin{equation}
\label{eq:LOSSoriginal}
\mathcal{L}_{LUT} = \mathcal{L}_{mse} + \lambda_{s}\mathcal{R}_{s} + \lambda_{m}\mathcal{R}_{m},
\end{equation}
where $\lambda_{s}$ and $\lambda_{m}$ are trade-off coefficients, we follow \cite{zeng2020learning} and set $\lambda_{s} = 1 \times 10^{-4}$, $\lambda_{m} = 10$. The MSE Loss $\mathcal{L}_{mse}$ ensures the content consistency between enhancing result and target photo. Smooth Regularization $\mathcal{R}_{s}$ and Monotonicity Regularization $\mathcal{R}_{m}$ are used to ensure the output values of the LUT are smoothed and monotonic, and the relative brightness and saturation of the input RGB values are maintained, ensuring natural enhancement results.

\textbf{Edge supervision loss. }
In the proposed local-context aware module, we predict the local attention mask $\mathbf{a}$ and employ the pixel-wise attention $a_{i,j}$ to modulate neighboring context pixels. We find apply supervision can improve the quality of local attention mask. As shown in \cref{bigfigurea}, given the human region mask of each photo, we can obtain the affinity of a pixel and its neighboring pixels via pair-wise inclusive-or operation. This affinity map can serve as the learning target for the local attention mask, and we guide the learning process by calculating the BCELoss. Given a mask, most pixels belong to the human region or background region, where the inclusive-or operation would generate an all-one mask. This would cause dramatically imbalanced loss values and impact the quality of the local attention mask. To alleviate this issue, we only calculate the loss on edge pixels, whose contextual pixels include both foreground and background, and name this loss edge supervision loss $\mathcal{L}_{edge}$.

\textbf{Group-style aware loss. }
The GAM-loss can be calculated as follows:
\begin{equation}
\label{eq:GLCloss}
\begin{aligned}
\mathcal{L}_{GAM} = \lambda_{GAM}(&Var(\mu_{I_{1}^a} \dots \mu_{I_{i}^a} \dots \mu_{I_{n}^a})\\
+&Var(\mu_{I_{1}^b} \dots \mu_{I_{i}^b} \dots \mu_{I_{n}^b})),
\end{aligned}
\end{equation}
where $\lambda_{GAM}$ is the coefficient of $\mathcal{L}_{GAM}$. In experiments, we find $\lambda_{GAM}$ = 0.001 is a proper choice. $\mu_{I_{n}^a}$ and $\mu_{I_{n}^b}$ indicates the average value of the $n^{th}$ input photo in a-channel and b-channel, respectively. $Var(\cdot)$ stands for calculating the variance of all elements.

In summary, the total loss is given by
\begin{equation}
\label{eq:loss}
\mathcal{L}  =  \mathcal{L}_{LUT} + \mathcal{L}_{edge} +  \mathcal{L}_{GAM}.
\end{equation}


\section{Experiments}

\begin{table*}[htbp]
 \centering
  \caption{Ablation experiments. (a) studies the influence of different spatial sizes when considering the local context. (b) reports the performance when varying the coefficient of group-style aware loss. (c) and (d) verify the influence of 3D LUT's number and the dimension in each LUT unit, respectively. In addition, we recorded the average running time (in milliseconds) of these models. All methods were tested on an NVIDIA 2080Ti GPU.}
    \begin{tabular}{c|cccccccc|ccccccc}
\cline{1-7}\cline{9-15} 
(a) & 
\multicolumn{1}{p{3.00em}}{\small $PSNR$} &
\multicolumn{1}{p{1.00em}}{\small $\Delta E_{ab}$} &
\multicolumn{1}{p{3.00em}}{\small $PSNR^{HC}$} &
\multicolumn{1}{p{1.00em}}{\small $\Delta E_{ab}^{HC}$} &
\multicolumn{1}{p{2.00em}}{\small $M_{GLC}$} &
\multicolumn{1}{p{1.50em}}{\small $Time$} &
&
(c) &
\multicolumn{1}{p{3.00em}}{\small $PSNR$} &
\multicolumn{1}{p{1.00em}}{\small $\Delta E_{ab}$} &
\multicolumn{1}{p{3.00em}}{\small $PSNR^{HC}$} &
\multicolumn{1}{p{1.00em}}{\small $\Delta E_{ab}^{HC}$} &
\multicolumn{1}{p{2.00em}}{\small $M_{GLC}$}&
\multicolumn{1}{p{1.50em}}{\small $Time$}\\
\cline{1-7}\cline{9-15}    
3 $\times$ 3  &\textbf{26.23} & \textbf{6.62} & \textbf{29.53} & \textbf{4.29} & 11.08 & \textbf{4.66}  &   & 3  & 25.91 & 6.84 & 29.18 & 4.44 & \textbf{10.84} & \textbf{4.56}\\
 5 $\times$ 5  & 26.19  & 6.67  & 29.50  & 4.32  & 11.39  & 4.80 &   & 5  & \textbf{26.23} & \textbf{6.62} & \textbf{29.53} & \textbf{4.29} & 11.08 & 4.66\\
7 $\times$ 7  & 26.05  & 6.73  & 29.33  & 4.36  & \textbf{11.06}  & 4.86  &   & 7  & 26.06  & 6.70  & 29.34  & 4.34  & 11.38 & 5.82\\
\cline{1-7}\cline{9-15} 
\multicolumn{1}{p{1.00em}}{}&       &       &       &       &       &  &     &\multicolumn{1}{p{1.00em}}{}&       &       &       &       &  \\
\cline{1-7}\cline{9-15} 
(b) &
\multicolumn{1}{p{3.00em}}{\small $PSNR$} &
\multicolumn{1}{p{1.00em}}{\small $\Delta E_{ab}$} &
\multicolumn{1}{p{3.00em}}{\small $PSNR^{HC}$} &
\multicolumn{1}{p{1.00em}}{\small $\Delta E_{ab}^{HC}$} &
\multicolumn{1}{p{2.00em}}{\small $M_{GLC}$} & 
\multicolumn{1}{p{1.50em}}{\small $Time$} &
&
(d) & \multicolumn{1}{p{3.00em}}{\small $PSNR$} &
\multicolumn{1}{p{1.00em}}{\small $\Delta E_{ab}$} &
\multicolumn{1}{p{3.00em}}{\small $PSNR^{HC}$} &
\multicolumn{1}{p{1.00em}}{\small $\Delta E_{ab}^{HC}$} &
\multicolumn{1}{p{2.00em}}{\small $M_{GLC}$}&
\multicolumn{1}{p{1.50em}}{\small $Time$}\\

\cline{1-7}\cline{9-15} 
1$\times10^{-2}$ & 24.78 & 8.47 & 28.10 & 5.48 & \textbf{5.99} & 4.66 & & 17 & 25.76 & 6.85 & 29.05 & 4.44 & \textbf{10.78} & \textbf{4.59}\\
  5$\times10^{-3}$ & 25.07 & 8.06 & 28.37 & 5.22 & 6.13 & 4.66 &  & 33  & \textbf{26.23} & \textbf{6.62} & \textbf{29.53} & \textbf{4.29} & 11.08 & 4.66\\
 1$\times10^{-3}$ & 25.66 & 7.42 & 28.96 & 4.82 & 6.67 & 4.66 &  & 49 & 26.12 & 6.72 & 29.41 & 4.36 & 11.51 & 4.70\\
1$\times10^{-4}$ & \textbf{26.05} & \textbf{6.99} & \textbf{29.34} & \textbf{4.52} & 7.87 & 4.66 & & 65  & 26.21  & 6.70  & 29.50  & 4.35  & 11.75 & 4.74\\
\cline{1-7}\cline{9-15} 
    \end{tabular}%
  \label{tab:Ablation_1}%
\end{table*}%

\begin{table}[htbp]
  \centering
  \caption{Ablation experiments on each component of our proposed module. 3D LUT was used as the baseline, and the effectiveness of each module is verified by step-to-step ablation.}
  \resizebox{0.49\textwidth}{!}{
    \setlength\tabcolsep{8pt}
    \renewcommand\arraystretch{1.0}
    \begin{tabular}{p{1.00em}ccccc}
    \toprule
        No.  & Model & $PSNR$ & $PSNR^{HC}$ & $M_{GLC}$ \\
    \hline
    \#1 & \multicolumn{1}{l}{3D LUT\cite{zeng2020learning}} & 25.99  & 28.29  & 10.81  \\
    \#2 & \multicolumn{1}{l}{\#1 + LAM} & 26.13  & 29.42  & 11.58  \\
    \#3 & \multicolumn{1}{l}{\#2 + Full Sup.} & 26.13  & 29.40  & 11.26  \\
    \#4 & \multicolumn{1}{l}{\#2 + Edge Sup.} & \textbf{26.23} & \textbf{29.53} & 11.08  \\
    \#5 & \multicolumn{1}{l}{\#4 + GAM}  & 25.66  & 28.96  & \textbf{6.67}\\
    \bottomrule
    \end{tabular}}%
    \vspace{-0.4cm}
  \label{tab:Ablation_2}%
\end{table}%

\subsection{Experiment setups}

\textbf{Datasets.}
All our experiments are performed on the PPR10K dataset \cite{liang2021ppr10k}. It provides 4K $\sim$ 8K high-resolution (HR) photos and corresponding 360P low-resolution (LR) photos with the retouched results by three experts. We use LR photos for training with randomly disrupted, while all of the LR and HR photos are available for testing. Following previous work \cite{liang2021ppr10k}, we divided the PPR10K dataset into a training set with 1,356 groups and 8,875 photos, and a testing set with 325 groups and 2,286 photos.

\textbf{Implementation details. }
We perform our experiments on RTX 2080Ti GPUs with Pytorch framework \cite{paszke2019pytorch}. 
We evaluate the performance of different settings and determine the size of the image patch is 3 $\times$ 3. 
All of the training photos are LR photos, so as to improve the training speed, and the testing photos include HR photos (4K $\sim$ 8K) and LR photos (360P). We follow the data augment setting in \cite{liang2021ppr10k} and train 200 epochs with batch size 16. We use Adam \cite{kingma2014adam} as the network optimizer and the learning rate is fixed at $1 \times 10^{-4}$ following \cite{zeng2020learning}.

\textbf{Evaluation metrics. }
Five metrics are employed to evaluate the performance of different methods quantitatively. Besides basic $PSNR$, the color difference between the retouched photo $R$ and the target photo $T$ is defined as the L2-distance in CIELAB color space with 
\begin{equation}
\label{eq:eab}
\Delta E_{ab} = \parallel R^{Lab} - T^{Lab}\parallel_{2}.
\end{equation}
The $PSNR^{HC}$ and $\Delta E_{ab}^{HC}$ mean human-centered PSNR and color difference. We follow the study of Liang \etal \cite{liang2021ppr10k} and adopt the same evaluation metric of group-style $\mathcal{M}_{GLC}$ to evaluate the performance of our model.

\subsection{Comparisons with state-of-the-art methods}

\cref{tab:bigtable} reports the comparison results with multiple state-of-the-art methods. The comparison results show that our proposed method has made obvious progress in both HR and LR photos. Using a specialist's retouching example, the PSNR of the HR photo is improved by 0.60, and the addition of the GAM led to a significant reduction in the $M_{GLC}$ metric, which demonstrates the effectiveness of our method.

\begin{figure*}[thbp]
\centering
\includegraphics[width=1\linewidth]{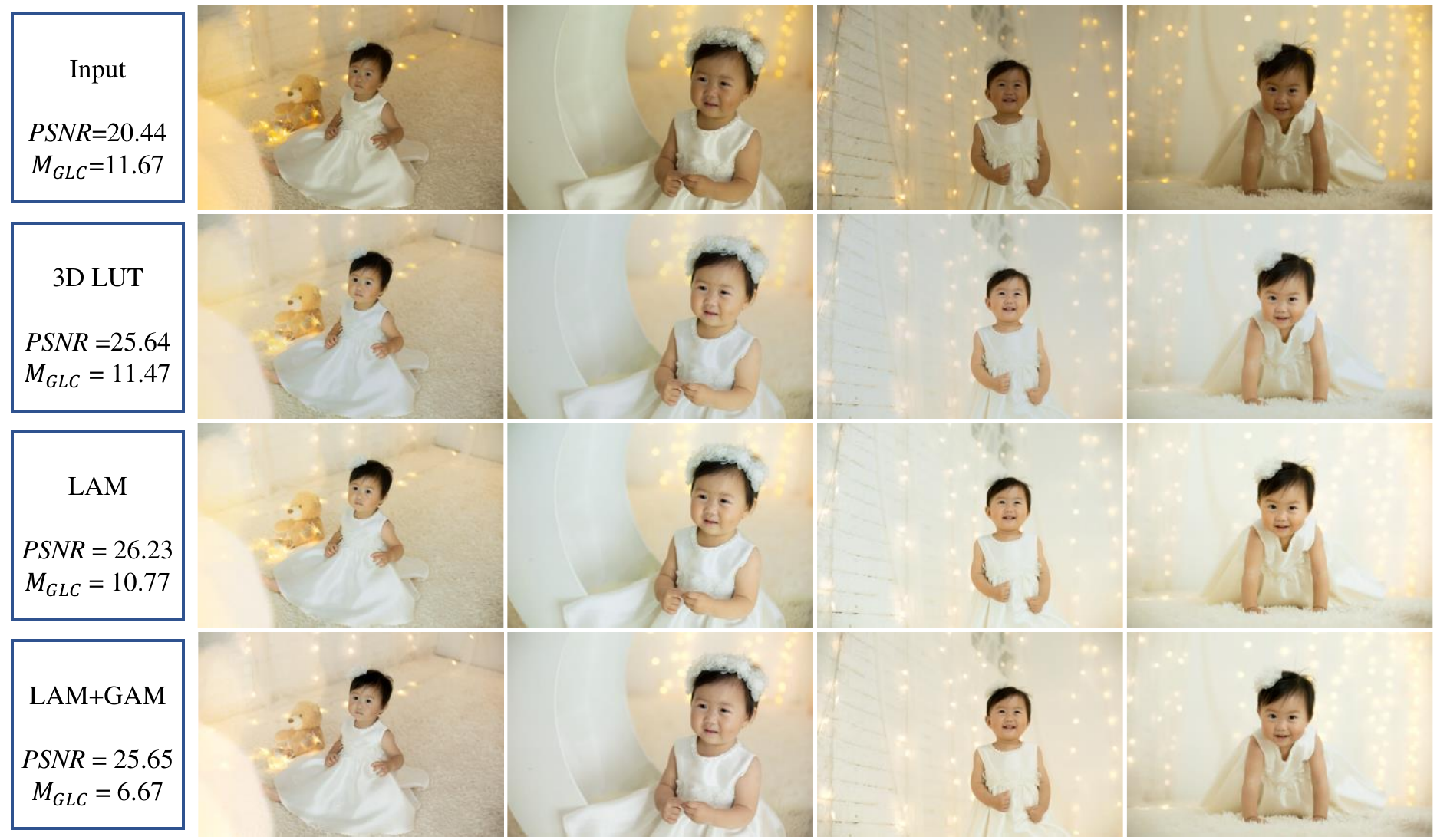}
\caption{Qualitative effect of the LAM and LAM+GAM strategy. Using the LAM improves the PSNR, and adding the GAM improves the tonal uniformity of a group of photos.}
\label{images}
\end{figure*}

The qualitative results are shown in \cref{images}, which not only demonstrate the photo retouching results of each network component but also show the input photo and target photo together. It can be intuitively seen that our method achieves the closest retouching results to the target photo. The results of retouching the same set of input photos with the local-context aware module are better than 3D LUT\cite{zeng2020learning}. While the output photos with the group-style aware module obviously maintain a more uniform retouching style, but the PSNR value only drops a bit. Meanwhile, the output of the photos with our proposed method has better scores.


\subsection{Ablation study}

We conducted extensive ablation experiments on the PPR10K dataset to determine the validity of the components and verify the influence of each parameter.

\textbf{Efficacy of each component. } 
We conduct extensive experiments to analyze the key components of our model, as shown in \cref{tab:Ablation_2}. We regard the model in \cref{sec:3D LUT} as the baseline model and add the local-context aware module on top of it. From the ablation experiments, we can find that all the metrics have been improved. After that, we design two different supervision strategies. Both of them are groundtruth mask oriented. They are full supervision corresponding to the whole photo and edge supervision for the attention mask. Both experiments are conducted on the basis of the local-contextual aware module. The results demonstrate that full supervision does not bring a significant improvement, almost any difference, while the edge supervision can consistently improve the performance.

\begin{table}[htbp]
  \centering
  \caption{Ablation experiments that replace the local-context aware module with UNet \cite{unet} to generate pixel-adaptive weights. }
    \begin{tabular}{ccccccc}
    \toprule
    \multicolumn{1}{p{2.00em}}{Model}&
    \multicolumn{1}{p{2.00em}}{\small $PSNR$} &
    \multicolumn{1}{p{1.00em}}{\small $\Delta E_{ab}$} &
    \multicolumn{1}{p{3.00em}}{\small $PSNR^{HC}$} &
    \multicolumn{1}{p{1.00em}}{\small $\Delta E_{ab}^{HC}$} &
    \multicolumn{1}{p{2.00em}}{\small $M_{GLC}$} & 
    \multicolumn{1}{p{2.00em}}{\small $Time$}\\
    \hline
    \multicolumn{1}{p{2.00em}}{\small LAM} & \textbf{26.23} & \textbf{6.62} & \textbf{29.53} & \textbf{4.29} & \textbf{11.08} & 4.66\\
    \multicolumn{1}{p{2.00em}}{\small UNet} & 25.72  & 6.92  & 28.97  & 4.49  & 11.15 & 5.88\\
    \bottomrule
    \end{tabular}
    \vspace{-0.4cm}
  \label{tab:LAMUNET}%
\end{table}%

We think the reason is due to the existence of the large pure background or pure foreground region in the groundtruth mask, which led to dramatically imbalanced supervision. Thus the generated loss is not accurate enough to learn. Edge supervision has a better learning efficiency because it focuses more on the demarcation of the photo. Finally, we add the group-style aware module and analyze that the previous strategy does not work well for group-level consistency, whose performance under the metric $M_{GLC}$ are greater than 10. However, after introducing our group-style aware module, the indicator can be reduced to 6.67. The substantial numerical improvement demonstrates the effectiveness of the proposed group-style aware module.

\textbf{Comparison to spatial-aware 3D LUT\cite{wang2021real}. } 
We follow spatial-aware 3DLUT \cite{wang2021real} and employ the same UNet architecture to replace our local-context aware module and predict pixel-adaptive LUT weights. As shown in \cref{tab:LAMUNET}, under the metric PSNR, our local-context aware module exceeds the UNet architecture with 0.51 points, and the group-level consistency and running time also verify the superiority of our method. We think there are two potential reasons. Firstly, UNet analyzes the contextual information for the whole image instead of a controlled receptive field, thus introducing some noise in the portrait photo retouching process. Secondly, the UNet architecture contains more parameters than our local attention mask module, which may cause the UNet architecture not easy to train.

\textbf{The parameter of 3D LUTs. } 
In order to ensure that the experiment settings are appropriate, we change the parameters for different experiments while keeping other conditions constant. Results of the experiments are shown in \cref{tab:Ablation_1}(c) and \cref{tab:Ablation_1}(d). The ablation results indicate that the most suitable number of 3D LUTs is 5, while the suitable dimension is 33.

\textbf{The size of the receptive field. } 
We consider that the processing of low-frequency information (brightness, color, \etc) in PPR tasks relies on a large receptive field, but a too-large receptive field may introduce extra noise and also lead to a larger amount of parameters in the convolutional block. To investigate the most suitable value for the patch (\eg receptive field) size of the LAM, we carry out experiments and report the results in \cref{tab:Ablation_1}(a). The experimental results indicate that the size of 3 $\times$ 3 is a proper choice for our local-attention aware module.

\textbf{Determining the coefficient of GAM.  } 
In the practical experiments, the introduction of the GAM module leads to a slight decrease in the retouching effect. Thus, it is unreasonable to purely optimize the group-level consistency and lead to large deviations in the retouching results. As shown in \cref{tab:Ablation_1}(b), we conduct ablation experiments and set the coefficient $\lambda_{GLC}$ to $1\times10^{-3}$. This makes a balance between the single-photo retouching quality (measured by PSNR) and group-level consistency (measured by $M_{GLC}$). 

By integrating the above experimental results, we ensure that parameters in our network are appropriate for the model. Due to the effectiveness of the local-context aware module and the group-style aware module, our proposed method achieves promising results. We are forming a concise and efficient solution for portrait photo retouching.

%
\section{Conclusion}
\vspace{-0.4cm}
As the existing LUT-based photo retouching paradigm \cite{zeng2020learning} only conducts pixel-to-pixel transformation but ignores the context cues, this paper designs a lightweight local-context aware module to incorporate context pixels and adaptively estimate LUT weights when tackling each pixel. Besides, a local attention mask is learned to distinguish the affinity of neighboring pixels further. Moreover, we propose to constrain the group-level consistency in the Lab space and guide the color temperatures and tones for a group of photos to be consistent. As we only employ four convolutional layers to capture local context cues explicitly, our method still maintains the lightweight superiority of the LUT-based photo retouching paradigm. Extensive experiments on the large-scale PPR10K dataset demonstrate the efficacy of the proposed local-context aware module and group-style aware module. However, as we only verify our method on a benchmark dataset, it may observe performance drop when tackling practical photos. In addition, the unintended usage of our method for surveillance may violate individual privacy. In the future, it is a promising direction to introduce the local-context aware module to other works, \eg super-resolution \cite{zhang2021context}, image classification  \cite{tang2015improving}, image inpainting \cite{zeng2019learning, zeng2021cr}.


{\small
\bibliographystyle{ieee_fullname}
\bibliography{egbib}
}

\end{document}